\documentclass[letterpaper,aps,prd,twocolumn,preprintnumbers,showkeys,nofootinbib,nobibnotes]{revtex4}
\usepackage{epsfig}
\usepackage{dcolumn}
\usepackage{bm}

\usepackage{graphicx}
\usepackage{latexsym}

\def\be{\begin{equation}}
\def\ee{\end{equation}}
\def\bea{\begin{eqnarray}}
\def\eea{\end{eqnarray}}
\def\map{{{WMAP} }}

\begin{document}

\title{Double Inflation and the Low CMB Quadrupole}
\author{Bo Feng}
\email{fengbo@mail.ihep.ac.cn}
\author{Xinmin Zhang}
\email{xmzhang@mail.ihep.ac.cn} \affiliation{Institute of High
Energy Physics, Chinese Academy of Sciences, P.O. Box 918-4,
Beijing 100039, People's Republic of China}

\begin{abstract}
Recent released WMAP data show a low value of quadrupole in the
CMB temperature fluctuations, which confirms the early
observations by COBE. In this paper we consider a model of two
inflatons with different masses,$~ V(\phi_1,\phi_2)={1\over
2}m_1^2\phi_1^2 + {1 \over 2} m_2^2 \phi_2^2~,$ $m_1>m_2$ and
study its effects on CMB of suppressing the primordial power
spectrum $P(k)$ at small $k$. Inflation is driven in this model
firstly by the heavier inflaton $\phi_1$, then the lighter field
$\phi_2$. But there is no interruption in between. We numerically
calculate the scalar and tensor power spectra with mode by mode
integrations, then fit the model to WMAP temperature correlations
TT and the TE temperature-polarization spectra. Our results show
that with $m_1\sim 10^{14}$ GeV and $m_2\sim 10^{13}$ GeV, this
model solves the problems of flatness {\it etc.} and the CMB
quadrupole predicted can be much lower than the standard power-law
$\Lambda$CDM model.
\end{abstract}


\maketitle Recently the Wilkinson Microwave Anisotropy Probe
(WMAP) data \cite{Bennett,Spergel,Verde,Peiris, Komatsu} have been
released and it is shown that the data is consistent with the
predictions of the standard ${\rm\Lambda}$CDM model with an almost
scale-invariant, adiabatic and Gaussian primordial (scalar)
fluctuations. However, there remain intriguing discrepancies
between the model and the observations, which show the
overprediction of the model on the amplitudes of fluctuations at
both the largest and the smallest scales. In Ref.\cite{Spergel}
Spergel {\it et al.} include other data of the Cosmic Microwave
Background(CMB)\cite{cbi,acbar}
 and Large Scale Structure(LSS)\cite{2df,forest}.
They find that for power law $\Lambda$CDM  model the best
 fit for the amplitude of fluctuations gradually drops as the
 probe of scale $k$ increases and the data supports for a nonzero running
of the scalar
 spectrum index from blue to red at $2\sigma$ with
 $dn_S/d \ln k = - 0.042^{+0.021}_{-0.020}$ \cite{Spergel}.
In Ref.\cite{seljak}, the authors have questioned about the
validity of the use of Lyman--$\alpha$ forest data \cite{forest},
despite this a slightly running of the power spectrum index is
still favored in the analysis of Refs.
\cite{Spergel,seljak,Lewis,Wang,Caldwell}.
 Especially, with a detailed reconstruction of the power spectrum Mukherjee and Wang\cite{Wang}
 have shown a preferred feature at $k\sim 0.01$
Mpc$^{-1}$, consistent with a running of the index. And in
Ref.\cite{Lewis} Bridle {\it et al.} have stressed the importance
of the data for the first three multipoles $l=2,3,4$ on the
requirement for the running index.

Theoretically there have been studies in the literature since the
release of the WMAP data on models of inflation which provide a
running index required by the WMAP
\cite{feng,lidsey,kawasaki,huang,CST}. However,
 as shown in Ref.\cite{Spergel}, the probability of finding a lower value of
quadruple in the presence of a {\bf constant} running of the
spectral index is no more than 0.9 percent for a spatial flat
$\Lambda$CDM cosmology. Indeed, the lack of CMB power on the large
angular scales  seen already in COBE \cite{cobe} \cite{Fang} and
reinforced by WMAP is more challenging.
 This discrepancy may be due to cosmic variance. On
the other hand, it probably gives a hint for new physics.

Recently several possibilities to $\bf{alleviate}$ the
low-multipoles problem have been proposed in the literature
\cite{Spergel,Tegmark,Efstathiou,Bridle,Lewis,Linde,Gaztanaga,Cline}
which include considering the effect of a finite
universe\cite{Spergel}, the late time integrated Sachs-Wolf (ISW)
effect by quintessence\cite{Bridle,Linde}, a non-flat
universe\cite{Efstathiou}, a suppression of the primordial
fluctuations\cite{Lewis,Linde,Cline} or different release of the
data\cite{Tegmark,Gaztanaga}. In the framework of inflation,
Contaldi {\it et al} in\cite{Linde} have discussed two approaches
to suppressing the large scale power. Beside the one of changing
the inflaton potential, they have proposed another one of changing
the initial conditions at the onset of inflation relative to the
standard chaotic inflation model\cite{chaotic}. For the latter
case, the inflaton has to be assumed in the kinetic dominated
regime initially.

In this paper we consider a double inflation model and study the
possibility of suppressing the lower multiples in the CMB. For a
quantitative investigation we study a model\cite{double}:
\be\label{potential} V(\phi_1,\phi_2)={1\over 2}m_1^2\phi_1^2 + {1
\over 2} m_2^2 \phi_2^2~. \ee The double inflation in the
literature has been studied widely. And phenomenologically the
model in (1) can be realized naturally in particle physics. For
example, in the sneutrino inflation models\cite{Yanagida,Ellis},
there are three sneutrinos which belong to three different
families. Taking two of them degenerated, it is effectively a
model of double inflation.

We assume in (1) that $\phi_1$ is heavier than $\phi_2$, {\it
i.e.} $m_1>m_2$. The inflation is firstly driven by $\phi_1$, then
by $\phi_2$, and there is no interruption in between. The
transition takes place at $m_{\phi_1}\sim H$, where $H$ is the
Hubble rate. When the transition happens $\phi_1$ starts to
oscillate around the minimum of its potential. We denote the
wavenumber of comoving mode which crosses the horizon around this
moment as $k_{f}$. Choosing the model parameters so that $k_{f}$
corresponds to the scale around our current horizon, we will show
in this paper that model (1) provides a scalar power spectrum much
suppressed around $k_{f}$. With a set of the model parameters we
will give a specific example of the initial power spectra and fit
the spectra to WMAP data. Our results show that the spectrum with
a feature is favored and lower CMB multipoles can be achieved by
the spectrum with a feature provided by this model.

For the discussions on double inflation, we use the notations of
Ref.\cite{Gordon}. In a spatially flat Friedmann-Robertson-Walker
(FRW) universe the evolution of the background fields for the
potential given in (1) is described by the Klein-Gordon equation:
\bea\label{eq:KG} \ddot{\phi}_I + 3H\dot{\phi}_I + V_{\phi_I}= 0~,
\eea and the Friedmann equation:
\begin{equation}
H^2=(\frac{\dot a}{a})^2 =\frac{8\pi G}{3} \left[
\frac{1}{2}\dot\phi_1^{~2}+ \frac{1}{2}\dot\phi_2^{~2} +V
\right]\,, \label{eq:hubble}
\end{equation}
where $I=1,2$, $a$ is the scale factor, the dot stands for time
derivative and $V_{x} = {\partial V}/{\partial x}$. Scalar linear
perturbations to the FRW metric can be expressed generally as (we
use the metric convention $+,-,-,-$): \be ds^2 =(1+2A)dt^2
-2aB_{,i}dx^idt - a^2[ (1-2\psi)\delta_{ij} + 2E_{,ij}] dx^idx^j .
\ee Thus the equation for the evolution of the perturbation
$\delta \phi_I$ with comoving wavenumber $k$ is given by
\begin{eqnarray}
\ddot{\delta\phi}_I + 3H\dot{\delta\phi}_I
 + \frac{k^2}{a^2} \delta\phi_I + \sum_J V_{\phi_I\phi_J}
\delta\phi_J= \nonumber\\ -2V_{\phi_I}A + \dot\phi_I \left[
\dot{A} + 3\dot{\psi} + \frac{k^2}{a^2} (a^2\dot{E}-aB) \right]
\,. \label{eq:perturbation}
\end{eqnarray}
Defining the adiabatic field $\sigma$ and its perturbation as
\cite{Gordon}: \bea \dot\sigma &=&(\cos\theta) \dot\phi_1 +
(\sin\theta) \dot\phi_2
\,,\nonumber\\
\delta\sigma &=& (\cos\theta) \delta\phi_1 + (\sin\theta)
\delta\phi_2\,, \eea with\footnote{Our definitions of $\cos\theta$
and $\sin\theta$ have the opposite signs with those in Ref.
\cite{Gordon}, but these differences do not affect the results.}
\begin{equation}
  \label{eq:cos sin}
\cos\theta = -\frac{\dot{\phi_1}}{\sqrt{\dot{\phi_1}^2 +
\dot{\phi_2}^2}}, \quad \sin\theta =
-\frac{\dot{\phi_2}}{\sqrt{\dot{\phi_1}^2 + \dot{\phi_2}^2}}\,.
\end{equation}
The background equations (3) and (2) become \bea H^2 &=&\frac{8\pi
G}{3} (\frac{1}{2}\dot\sigma^{~2}+V
)\,,\nonumber\\
\ddot{\sigma} &+& 3H\dot{\sigma} + V_\sigma = 0\,, \eea where
$V_\sigma=(\cos \theta) V_{\phi_1} + (\sin\theta) V_{\phi_2}$. The
comoving curvature perturbation is given by \cite{Gordon}
\begin{eqnarray}
\label{eq:zeta} {\cal R}= \psi + \frac{H}{\dot{\sigma}}
\delta\sigma \,.
\end{eqnarray}
We assume that there is no entropy perturbation, and this is
consistent with the results of WMAP \cite{Peiris}. Under this
assumption, there is an adiabatic condition between $\delta
\phi_1$ and $\delta \phi_2$:
\begin{equation}
\label{eq:adiabatic}
 {\delta\phi_1\over\dot{\phi_1}}={\delta
\phi_2\over\dot{\phi_2}}~.
\end{equation}
So, the equation governing the evolution of adiabatic perturbation
is the same as that in the single field inflation model
\cite{mfb,SL}:
\begin{equation}\label{equ}
u''+ (k^{2}-\frac{z''}{z} )u=0~,
\end{equation}
where $u=-z{\cal R}$ and $z=a\dot\sigma/H$, the prime denotes the
derivative with respect to conformal time $\eta$ ($\eta=\int
\frac{dt}{a}$). The power spectrum of adiabatic perturbation is
defined as
\begin{equation}
P_{\cal R}(k) = \frac{k^{3}}{2\pi^{2}}|\frac{u}{z}|^2~,
\end{equation}
which approaches a constant at late time as $k/aH \rightarrow 0$.
Similarly for  tensor perturbations the power spectrum is
\begin{equation}
P_{g}(k) = \frac{k^{3}}{2\pi^{2}}|\frac{v}{z}|^2~,
\end{equation}
where $v= a \Psi$, $\Psi$ is the linear tensor perturbation
\cite{mfb,SL} and the equation of motion for $v$ is :
\begin{equation}\label{eqv}
v''+ (k^{2}-\frac{a''}{a} )v=0~.
\end{equation}

\begin{figure}
\includegraphics[scale=.4]{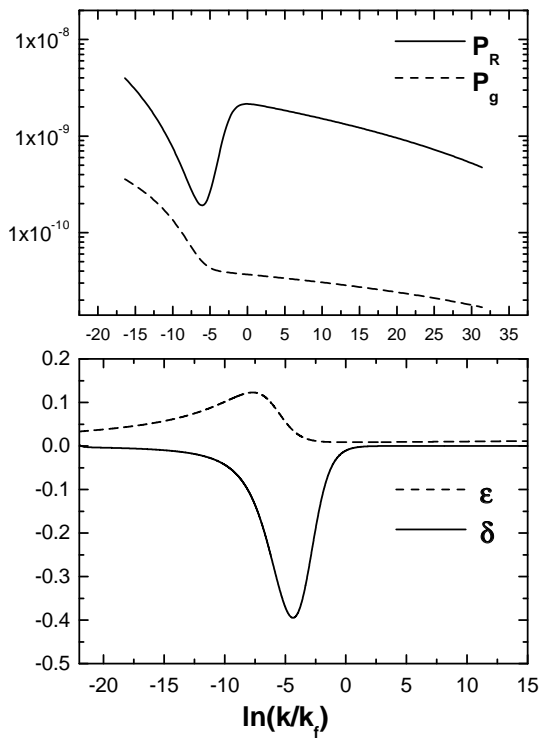}
\caption{Initial power spectra and slow rolling parameters as a
function of $\ln(k/k_f)$. $N(k_f)=59.6$}\label{fig:fig2}
\end{figure}

Using formulations above we are able to calculate the primordial
 power spectra with mode by mode
integrations\cite{wenbin,feng,wangxl}. Regarding the choices of
the model parameters: initial values of $\phi_1$ and $\phi_2$,
$m_1$ and $m_2$, we notice that $\phi_1$ is arbitrary with a weak
prior to provide enough number of $e-folding$ to solve the
flatness problem\cite{Linde,Cline}, $\phi_2$ mainly determines
which $P_R$ correspond to the cosmological scale and the ratio of
$m_1$ to $m_2$ determines the shape of $P_R$ with the absolute
value of $m_2$ fixed by the WMAP normalization. For different
values of $\phi_2$, the corresponding number of $e-folding$ to CMB
scales will differ and the shape of $P_R$ will also get changed,
as in the case of the single field inflation. The amplitude of
$P_R(k)$ is to be determined by observations. In
Fig.\ref{fig:fig2} we show initial power spectra as a function of
$\ln (k/k_f)$. In the numerical calculation\cite{note} we have set
$m_1=8 m_2$ and $\phi_2=3.3 M_{Pl}$ at the onset of inflation.
This gives rise to $N(k_f)=59.6$. We find such a value of $N(k_f)$
is acceptable for fitting to WMAP data below. We also show in
Fig.\ref{fig:fig2} the behavior of the slow rolling(SR) parameters
defined by $\epsilon \equiv -\dot H/H^2 $ and $\delta\equiv \ddot
\sigma/H \dot \sigma$~~. One can see that these parameters change
dramatically -- this is why we use numerical calculations instead
of the Stewart-Lyth analytical
formula\cite{SL,wenbin,Leach:2000yw}.

Now we fit the WMAP data with the primordial spectra in
Fig.\ref{fig:fig2}. Our modified version of the publicly available
CMBFAST \cite{cmbfast} is based on Version 4.2 \cite{IEcmbfast}
and we have used the "HP" choice to give exact CMB TT and TE power
spectra. We also run with CAMB\cite{camb,IEcamb} for a crosscheck
on our results. We use a similar method to Ref.\cite{Linde}, and
set $\Omega_{\Lambda}$ and $\ln k_f$ as free parameters in our
fit. Denoting $k_c=7.0\times 70./3/10^5\approx 1.6\times 10^{-3}$
Mpc$^{-1} $, we use 101 and 251 grid points with ranges
$[0.68,0.77]$, and $[-19.6,5.4]$ respectively for
$\Omega_{\Lambda}$ and $\ln (k_f/k_c)$. At each point in the grid
we use subroutines derived from those made available by the \map\
team to evaluate the log likelihood with respect to the \map\ TT
and TE data \cite{Verde}. Other parameters are fixed at
$\Omega_bh^2=0.022$, $\Omega_{m}h^2 = 0.135$ and $\tau_c=0.17$ and
$\Omega_{\rm tot}=1$ \cite{Spergel}. The overall amplitude of the
primordial perturbations has been used as a continuous parameter.
Differing from Ref.\cite{Linde} and Ref.\cite{Cline}, we have
included the tensor contributions in our fit to CMB. And for
comparison, we also run the code without tensor.

\begin{figure}
\includegraphics[scale=.3]{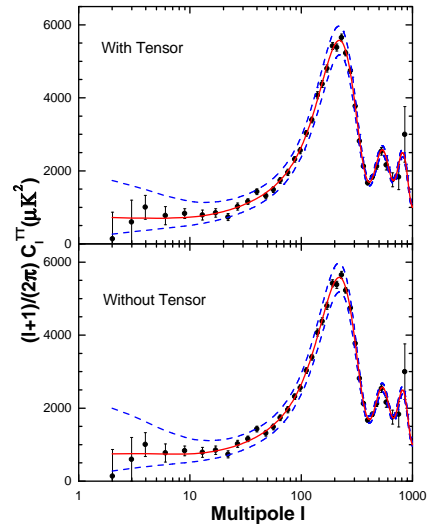}
\caption{CMB anisotropies with primordial spectra shown in
Fig.\ref{fig:fig2} . The error bars are taken the same as
Ref.\cite{Spergel}. The upper and lower dashed lines show the
1-$\sigma$ confidence levels for lognormal distributions with
cosmic variance limits. The middle solid lines show the models
with the lowest quadrupoles. }\label{fig:835}
\end{figure}

\begin{figure}
\includegraphics[scale=.3]{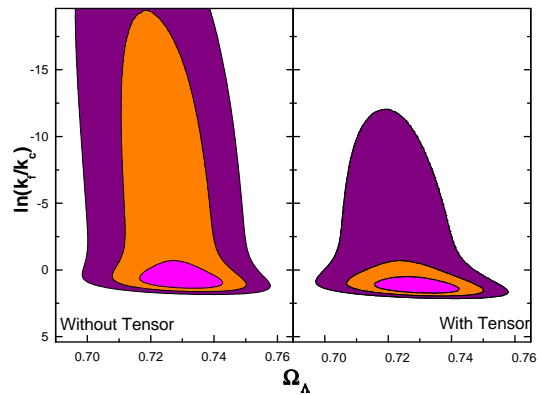}
\caption{ Two-dimesional contours in the $\ln(k_f/k_c)$--
$\Omega_{\Lambda}$ plane for our grids of model. $k_c \approx 1.6
\times 10^{-3}$ Mpc$^{-1} $. The regions of different color show
68.3\%, 95\% and 99.7\% confidence respectively.}\label{fig:2d}
\end{figure}

\begin{figure}[t]
\includegraphics[scale=.3]{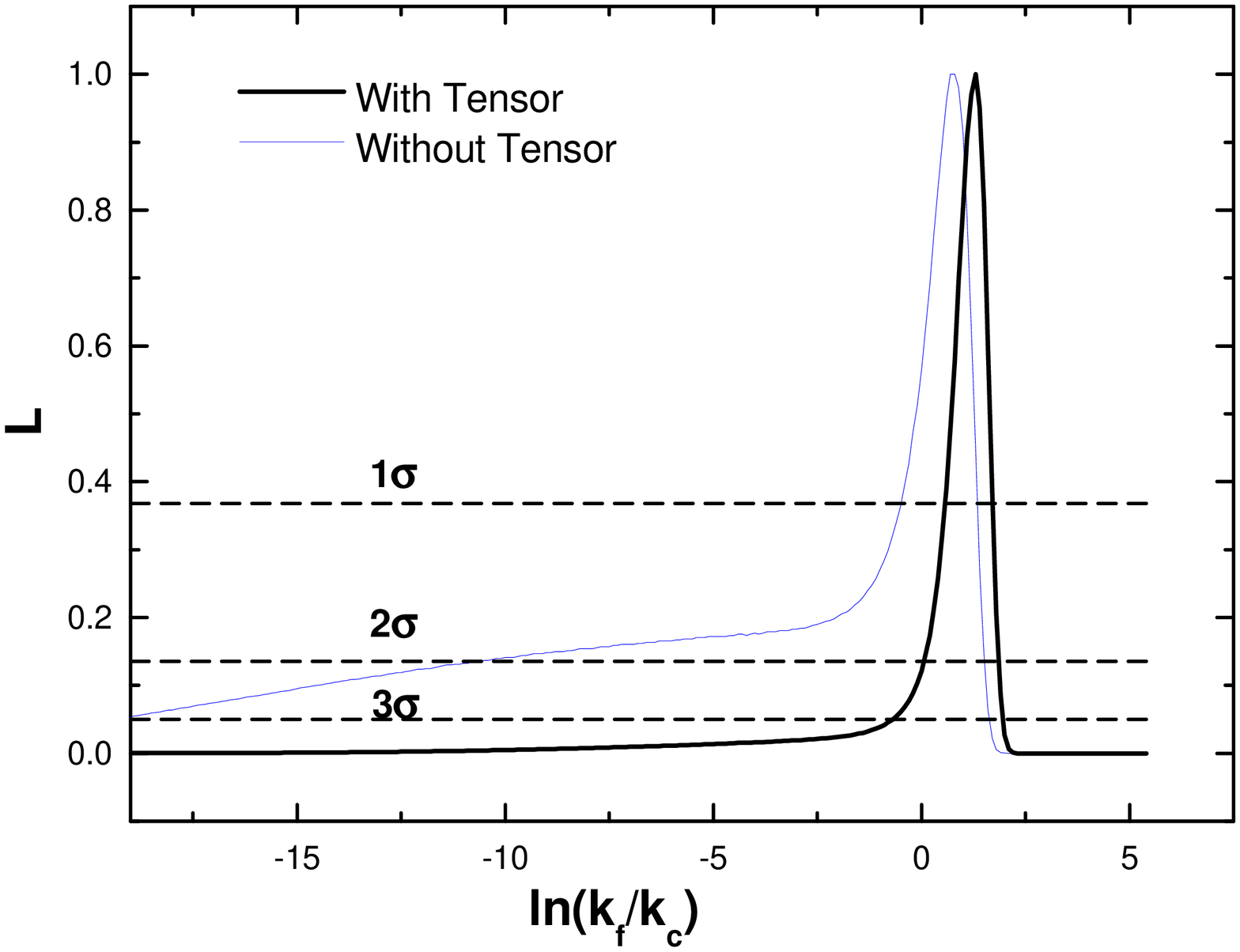} \caption{One-dimesional marginalized distributions for
the values of $\ln(k_f/k_c)$.}\label{fig:1d}
\end{figure}

In Fig.~\ref{fig:835} we show the resulting best--fit models
obtained from the grids for the model considered. In the plots we
take the same error bars on the binned WMAP results as
Ref.\cite{Spergel}. The regions between the two dashed lines are
given by 1-$\sigma$ confidence levels for  lognormal distributions
as well as the cosmic variance limits. The cosmic variance factor
is $1 \pm \sqrt{2/(2 l +1)} $. The middle solid lines show the
models with the lowest quadrupoles. We get the minimum
$\chi^2=1429.2$ when not including tensor and a slightly larger
$\chi^2=1429.7$ with tensor for the best fit values. One can see
the CMB low multipoles do have been suppressed in our model.

In Fig.~\ref{fig:2d} we plot the resulting $\chi^2$ values as
functions of $\Omega_{\Lambda}$ and   $\ln(k_f/k_c)$. The contours
shown are for $\Delta\chi^2$ values giving one, about two, and
three $\sigma$ contours for two parameter Gaussian distributions.
We find that the primordial spectrum with a feature is favored at
more than 3-$\sigma$ level. However, when neglecting the tensor
contribution, the significance reduces to only about 2-$\sigma$
level. In our model the ratio of tensor to
 scalar $r\equiv 8 P_g/P_R$ reaches its minimum at $N \sim N(k_f)$
and it gradually increases for smaller $N$ (larger $k$) and grows
rapidly for larger $N$ (smaller $k$). And our $N(k_f)$ corresponds
to the region with the lowest value of $r$, which is consistent
with WMAP group's analysis that large tensor contribution is
disfavored by current CMB observations\cite{Spergel,Peiris}. The
1-$\sigma$ regions in Fig.~\ref{fig:fig2} and Fig.~\ref{fig:2d}
differ slightly when with and without tensor since $r$ is around
its minimum in both cases.

We marginalize over $\Omega_{\Lambda}$ to obtain the
one-dimensional probability distributions in $\ln(k_f/k_c)$ shown
in Fig.~\ref{fig:1d}. For the spectra in Fig.\ref{fig:fig2} when
neglecting tensor contributions, we get $\ln(k_f/k_c)=0.8$ with
the maximum likelihood, corresponding to $k_f=3.6 \times 10^{-3}$
Mpc$^{-1}$. We also have $k_f\sim 0$ at $2\sigma$ level. When
taking into account tensor contributions, the maximum likelihood
value of $\ln(k_f/k_c)$ shift to 1.3 and we get $ \ln(k_f/k_c)=
1.3^{+0.4}_{-0.7} $    at $1\sigma$, $ 1.3^{+0.6}_{-1.2} $ at
$2\sigma$ and $ 1.3^{+0.7}_{-2.0} $ at $3\sigma$, corresponding to
$k_f=6.0^{+3.0}_{-3.1}\times 10^{-3}$ Mpc$^{-1}$,
$6.0^{+4.5}_{-4.3}\times 10^{-3} $Mpc$^{-1}$ and
$6.0^{+5.7}_{-5.2}\times 10^{-3} $Mpc$^{-1}$ respectively. The
difference in $\chi^2$ between the peak in the distributions and
at $\ln(k_f/k_c)= -19.6$ is found to be $\Delta\chi^2=6.1$ when
not including tensor and $\Delta\chi^2=18.3$ when with tensor.

 When considering tensor contributions in the two dimensional contour
  between $\ln(k_f/k_c)$ and the
 normalized factor of the primordial scalar spectrum at $k=0.05$
 Mpc$^{-1}$, we obtain $\ln(k_f/k_c)=
-0.8\sim 1.9 $  and  $P_R(0.05/Mpc)=2.48 \sim 2.53 \times 10^{-9}
$ at $2\sigma$ level and we get $m_1=1.3\sim 1.4 \times 10^{14}$
GeV. In contrast to the power law primordial spectrum with
constant $n_S$ we run a similar code: we fix $\Omega_bh^2=0.022$,
$\Omega_{m}h^2 = 0.135$ and $\tau_c=0.17$ and $\Omega_{\rm
tot}=1$, varying $\Omega_{\Lambda}$ and $n_S$ with ranges
[0.68,0.77], [0.91,1.07] and get a minimum $\chi^2=1432.7$. To
characterize the pure power law primordial spectrum, one considers
two parameters: $n_S$ and the amplitude. For our double inflation
model four parameters are introduced to give the exact scalar and
tensor spectra: $m_1$, $m_1/m_2$, $N(k_f)$ (or equivalently
$\phi_2$ at the onset of inflation) and $\ln(k_f/k_c)$. This
indicates our double inflation model is favored at $\sim 1.2
\sigma$ compared with power law $\Lambda$CDM model\cite{note2}. In
general, primordial power spectra with a feature or cutoff do
generate a lower CMB TT quadrupole (which, however may not be
sufficient)\cite{Lewis,Linde,Cline}. A cutoff primordial spectrum
also as pointed out in Ref.\cite{Cline} makes the CMB TE
multipoles lower. When combining these two effects, however, our
calculations show that the primordial spectrum with a feature can
work better than models with power law primordial spectra.

In conclusion, we have studied the possibility of suppressing the
low multipoles in the CMB anisotropy with a model of double
inflation. Our results show that with $m_1\sim 10^{14}$ GeV and
$m_2\sim 10^{13}$ GeV which lies in the parameter space required
by neutrino physics in the scenario of sneutrino
inflation\cite{Yanagida,Ellis}, this model fits to the WMAP data
better than the standard power-law $\Lambda$CDM model.

We thank Dr. Mingzhe Li for discussions and Dr. H. H. Peiris for
communications on WMAP data. We also thank the anonymous referee
for comments and suggestions. This work was supported in part by
National Natural Science Foundation of China and by Ministry of
Science and Technology of China under Grant No. NKBRSF G19990754.

\newcommand\AJ[3]{~Astron. J.{\bf ~#1}, #2~(#3)}
\newcommand\APJ[3]{~Astrophys. J.{\bf ~#1}, #2~ (#3)}
\newcommand\APJL[3]{~Astrophys. J. Lett. {\bf ~#1}, L#2~(#3)}
\newcommand\APP[3]{~Astropart. Phys. {\bf ~#1}, #2~(#3)}
\newcommand\CQG[3]{~Class. Quant. Grav.{\bf ~#1}, #2~(#3)}
\newcommand\JETPL[3]{~JETP. Lett.{\bf ~#1}, #2~(#3)}
\newcommand\MNRAS[3]{~Mon. Not. R. Astron. Soc.{\bf ~#1}, #2~(#3)}
\newcommand\MPLA[3]{~Mod. Phys. Lett. A{\bf ~#1}, #2~(#3)}
\newcommand\NAT[3]{~Nature{\bf ~#1}, #2~(#3)}
\newcommand\NPB[3]{~Nucl. Phys. B{\bf ~#1}, #2~(#3)}
\newcommand\PLB[3]{~Phys. Lett. B{\bf ~#1}, #2~(#3)}
\newcommand\PR[3]{~Phys. Rev.{\bf ~#1}, #2~(#3)}
\newcommand\PRL[3]{~Phys. Rev. Lett.{\bf ~#1}, #2~(#3)}
\newcommand\PRD[3]{~Phys. Rev. D{\bf ~#1}, #2~(#3)}
\newcommand\PROG[3]{~Prog. Theor. Phys.{\bf ~#1}, #2~(#3)}
\newcommand\PRPT[3]{~Phys.Rept.{\bf ~#1}, #2~(#3)}
\newcommand\RMP[3]{~Rev. Mod. Phys.{\bf ~#1}, #2~(#3)}
\newcommand\SCI[3]{~Science{\bf ~#1}, #2~(#3)}
\newcommand\SAL[3]{~Sov. Astron. Lett{\bf ~#1}, #2~(#3)}


\begin{thebibliography}{999}
\bibitem{note}
Since we have to resort to mode by mode integrations on $P_R$ and
fit it to WMAP data, it is time consuming to make a full search in
the model's parameter space. We leave it for future
investigations.
\bibitem{note2}
 When global fits are applied to the comparison, e.g., all
the background parameters vary, LSS and other CMB data are
considered, the $1.2 \sigma$ might get slightly
changed\cite{Cline}.
\end{thebibliography}
\end{document}